\relax
\documentstyle[12pt]{article}
\advance\textheight by \topskip
\begin{document}
%\bf
\def\W {{\cal W}}
\def\w {{  w }}

\begin{flushright}
QMW-PH-94-8\\
hep-th/9404073

\end{flushright}
\vspace{-0.2cm}
\begin{center}
{\large\bf ON $W_{ \infty/2}$ STRINGS}\\
\vskip 2 cm
{\bf CHRISTOPHER  HULL\\
\vspace{1ex}
and \\ ZOHORA KHATUN \\}

\vskip 0.6 cm
Department of Physics, Queen Mary and Westfield College,\\
\vskip 0.4 cm
Mile End Road,
London E1  4NS,
England.
\vskip 0.7 cm
\end{center}
\vskip 1.5 cm
\centerline{\bf ABSTRACT}
\begin{quotation}

The string theory defined by gauging   $\W_{ \infty/2}$, the sub-algebra of
$\W_{ \infty}$
generated by the currents of even spin, is discussed.  The critical value of
the central charge is calculated using zeta-function techniques and shown to be
$c=1$.  A critical string theory is constructed  by coupling a free boson to
$\W_{ \infty/2}$-gravity   and the physical states are analysed.

\end{quotation}

%\normalsize

\newpage

\def\rt {\rightarrow}
\def\+-{\underline +}
\newcommand{\beq} {\begin{equation}}
\newcommand{\eeq}{\end{equation}}
\newcommand{\bear}{\begin{eqnarray}}
\newcommand{\enar}{\end{eqnarray}}
\newcommand{\nmb}{\nonumber}

\def\np{Nucl.\ Phys. }
\def\pl{Phys.\ Lett. }
\def\pr{Phys.\ Rev. }
\def\prl{Phys.\ Rev.\ Lett. }
\def\cmp{Comm.\ Math.\ Phys. }
\def\intmp{Intern.\ J.\ Mod.\ Phys. }
\def\mpl{Mod.\ Phys.\ Lett. }

Any matter system with
a  $\W$ -algebra symmetry can be coupled, at least classically, to the
corresponding \W -gravity to obtain a theory with local $\W$-symmetry (a
higher-spin generalisation of diffeomorphism symmetry).
One can then attempt to quantize the theory, and a consistent critical
$\W$-string theory emerges provided
that the matter system is chosen so that all anomalies are cancelled by ghost
contributions (for a review, see e.g. \cite{rev}).  This is equivalent to the
existence of a nilpotent BRST charge, and requires in particular that the sum
of the matter and ghost central charges should be zero.
We will be concerned here with algebras such as $\W_\infty$ or
$\W_{1+\infty}$
that are generated by an infinite number of currents, and in particular in the
algebra
 $\W_{ \infty/2}$, which is the sub-algebra of  $\W_\infty$ generated by the
currents of even spin.
For such infinitely-generated $\W$-algebras, there are an infinite number of
ghosts and the ghost contribution to the central charge is the sum of a
divergent series. One approach to this problem is to regulate the series using
the  zeta-function method, giving a finite critical central charge. For
$\W_\infty$ this gives a critical value of the matter central charge given by
$c=-2$, and critical $\W_\infty$-string theories have been investigated in
\cite{ber}.
Remarkably, for the $\W_{ \infty/2}$ sub-algebra, the critical central charge
is $c=1$, so that a critical $\W_{ \infty/2}$-string can be constructed by
coupling a free boson to $\W_{ \infty/2}$-gravity, with a linearly-realised
$\W$-symmetry, so that there are no matter-dependent anomalies.
This is particularly interesting because of   recent developments in the study
of $c=1$ matter coupled to ordinary gravity, a system which  has been shown to
posses a $\W_\infty$ symmetry.
 It is the purpose of this paper to investigate the critical $\W_{
\infty/2}$-string theory.

The $w_\infty$ algebra \cite {baks} is generated by an infinite set of
currents $w^s, s=2,3........$, of spin s. The modes $w^s_m  (m\in Z)$
satisfy the $w_\infty$ algebra
\beq
[w^s_m, w^r_n]=[(r+1)m-(s+1)n] w^{s+r}_{m+n}
\eeq
which is a Lie algebra without central extension.
The $\W_\infty$ algebra \cite {pope} is a deformation of this which includes
the Virasoro algebra with central charge $c$, given by
\beq
[\W^r_n,\W^s_m]=\sum_{q\geq 0 }g^{rs}_{2q}(m,n) \W^{r+s-2q}_{m+n}
+c_r(m)\delta^{rs}\delta_{m+n,0}
\eeq
The structure constants $g^{rs}_{2q}(m,n)$ and the central terms
$c_r(m)$ are completely fixed by the Jacobi identities and take
the following form
\beq
	c_r(m)=m(m^2-1)(m^2-4)........(m^2-{(r+1)}^2)c_r
\eeq
where the central charges $c_r$ are given by
\beq
       c_r=\frac {2^{2r-3}r!(r+2)!}{(2r+1)!!(2r+3)!!}c
\eeq
The structure constants are given by
\beq
      g^{rs}_q(m,n)=\frac{1}{2(q+1)!}\phi^{rs}_q N^{rs}_q(m,n)
\eeq
where $N^{rs}_q$ is defined by
\beq
      N^{rs}_q(m,n)=\sum_{k=0}^{q+1} (-1)^k {q+1\choose k}
       [r+1+m]_{q+1-k} [r+1-m]_k [s+1+n]_k [s+1-n]_{q+1-k}
\eeq
with $[x]_n= \Gamma (x+1)/\Gamma (x+1-n)$.
Finally the $\phi^{rs}_q$ are given by
\beq
 \phi^{rs}_q =\sum_{k\geq0}
 \frac {{(-\frac{1}{2})}_k  {(\frac{3}{2})}_k {(-\frac{q}{2}- \frac{1}{2})}_k
  {(-\frac{q}{2})}_k }
  {k!{(-r- \frac{1}{2})}_k  {(-s- \frac{1}{2})}_k
  {(r+s- q+ \frac {5}{2})}_k }
\eeq
where ${(x)}^n=\Gamma(x+n)/\Gamma(x)$

   Restricting to the subset of currents of even spin, $w^{(2r)}$ or
   $\W^{(2r)}$, gives a
closed subalgebra of $w_\infty$ or $\W_\infty$ denoted $w_{\infty/2}$
or $\W_{\infty/2}$ respectively \cite{pope,he}.
Similarly restricting to the subset of currents  $w^{s},s=2,2+M,2+2M,2+3M...$
gives a subalgebra of $w_\infty$ denoted by $w_{{\infty}/M}$
\cite{he}. These do not appear to give a subalgebra of $ \W_\infty$, however,
and
it is not known whether a consistent non-degenerate algebra with non trival
central extention exists for currents with these spins if $M\not =2$.

   Given a classical matter system with $\W$-algebra symmetry, it is
straightforward to couple to $\W$-gravity by introducing a gauge field of
spin $s$ corresponding to each conserved $\W$-current of spin $s$ \cite{he}.
   On quantisation, one gauges away each gauge field of spin $s$ and
introduces a ghost  $c^{(s)}$ of spin $1-s$ and an anti-ghost $b^{(s)}$ of spin
$s$. In general,
there will be anomalies of two types \cite{hee}: (i) if the $\W$-currents do
not generate a closed quantum algebra, there will be matter-dependent
anomalies (ii) there will in general be universal anomalies, corresponding
to central terms in the current algebra. In particular, there will be a
conformal (or gravitational) anomaly corresponding to the total
central charge of the Virasoro algebra, given by
\bear
       c_{TOT}  &=& c_{MATTER}+c_{GHOST} \nmb \\
	    c_{GHOST} &=& \sum_{s=2}^\infty c_s \\
	    c_s &=& -2(6s^2-6s+1)\          \nmb
\enar
	This anomaly will be absent only if the matter contribution cancels
against the ghost contribution.

The requirement of anomaly cancellation is equivalent to the existence
of a nilpotent BRST charge $Q$. If the $\W$-algebra is a Lie algebra (as in
the case for algebras considered above) then it is straightforward to
construct a BRST charge by standard techniques and check its nilpotence.
A necessary condition for this is the vanishing of $c_{TOT}$.

The case of the $\W_\infty$ algebra was considerd in \cite{pape,yam}.
The ghost central charge is
\bear
      c_{GHOST} &=&\sum_{s=2}^\infty c_s \nmb \\
		&=& -26-74-146-........... \
\enar
which is divergent. However, it was suggested in \cite{pape,yam} that
this divergent sum can be regularised using a zeta-function techniques.
Recall that the zeta-function is
\beq
    \zeta(s)=\sum_{n=1}^\infty n^{-s}
\eeq
so that formally one can define $\sum _{n=1} ^\infty n^r$ (with $r\geq 0$)
to be $\lim_{s\rt -r} \zeta(s)$,  which is a well-defined finite number.
Thus
\beq
	c_{GHOST}\rightarrow-2[6\zeta(-2)-6\zeta(-1)+\zeta(0)]
\eeq
Using
\beq
	    \zeta(0)=-{1\over2},\quad \zeta(-1)=-{1\over 12},
	    \quad \zeta(-2)=0
\eeq
one finds the regularised value
\beq
			 c_{GHOST}=2
\eeq
So the anomalies cancel if the matter sector has central charge $c=-2$.

In \cite{pape} arguments were given to support this rather formal procedure.
It was conjectured that this condition was sufficient for the existence
of a nilpotent BRST charge and this was checked explicitly for a number
of higher spin anomalies.

Consider now the application of these ideas to $\W_{\infty/ 2}$. Then
\bear
	    c_{GHOST} &=&\sum_{r=1}^\infty c_{2r} \nmb\\
	      &=&-2\sum_{r=1}^\infty \left[ 6(2r)^2-6(2r)+1  \right]\\
	      &=&-2\{24\zeta(-2)- 12\zeta(-1) +\zeta(-1)\} \nmb
\enar
giving the regularised value $c_{GHOST}=-1$.

Similarly one can ask for the critical value of the central charge for
a theory of $\W_{\infty/M}$ gravity.
Then zeta-function regularisation gives
\bear
    c_{GHOST} & =&\sum_{r=0}^\infty c_{2+rM} \nmb \\
		  &=&(4M^2 -15M +13)\
\enar
Thus if a theory of critical $\W_{\infty/M}$ strings is to exist, the matter
sector must have central charge
	   $$ c_{MATTER}(M)=-(4M^2 -15M +13) $$
Note that $c_{MAT}<0 $    unless $M=2$, so that unless $M=2$, one might
expect problems with unitarity.

       In the case of $\W_\infty$, $M=1$,  $c_{MAT}=-2$,
a non-linear realisation of
       $\W_\infty$ with $c_{MAT}=-2$
in terms of one boson was constructed in \cite{pape} and used to construct a
$\W_\infty$-gravity theory that was formally anomaly-free.
       The fact that the $\W$-symmetry was non-linear meant that there were
matter-dependent anomalies \cite{hee} and the cancellation of these at $c=-2$
was
rather involved \cite{ber}. Unlike a $\W$-algebra generated by a finite
number of
currents (e.g. $\W_N$), $\W_\infty$ also  has linear realisations in
terms of free bosons.
Given $D$ free complex bosons $\phi^i,\bar\phi^i$ the set of currents
\beq
    \W^s=B(s)\sum_{r=1}^{s-1}(-1)^r A_r^s:\partial_z^r\phi
 \partial_z^{s-r}\bar\phi:           \label{shumy}
\eeq
where,
\beq
   B(s)=q^{s-2}{{2^{s-3}s!}\over (2s-3)!!}
\eeq
and
\beq
  A_r^s ={1\over(s-1)}{s-1\choose r}{s-1\choose s-r}
\eeq
satisfy the $\W_\infty$ algebra with $c=2D$ \cite{kiri}.

In the case of the $N$-free boson realisation of the Virasoro algebra
with $c=N$, it is possible to add a background charge term
$\alpha_i \partial^2 \phi^i$  to the stress
tensor so that  the central charge can take general values $c=N+24 \alpha^2$.
We have checked that it is not possible to modify the linear bosonic
realisation of $\W_\infty$ to obtain a realisation with
general central charges by adding  higher derivative terms  that are linear in
the bosonic fields to the currents (\ref{shumy}).

If one restricts the $D$ bosons $\phi^i$ to be real in this construction,
then the odd-spin currents given by (\ref{shumy}) vanish, $\W^{(2r+1)}=0$,
and one is left with a set of even spin currents generating the
$\W_{\infty/2}$ algebra with $c_{MAT}=D$.

       In particular, for one boson, $D=1$, and we have a realisation of
$\W_{\infty/2}$ with the critical $c$-value, $c_{MAT}=1$. This means
that with a matter system
of one boson we have a $\W$-gravity theory  for which the gravitational or
conformal anomalies cancel against ghost contributions and we conjecture
that the same is true for all higher spin anomalies as well. If this is so
then we have constructed a theory of critical $\W_{\infty/2}$-strings.

The operator
quantization of this free boson theory is straightforward. On the
cylindrical world-sheets with time coordinate $\tau$ and periodic
space-cordinate $\sigma$,
\beq
     \phi=\phi_+ (\tau+\sigma) + \phi_- (\tau-\sigma)
\eeq
and  \[ \partial \phi_+(\sigma) =\sum a_n e^{in\sigma}\]
where the modes $a_n$ satisfy
\beq
       [a_n,a_m]=n\delta_{n+m}
\eeq
 As usual, we choose a Fock space represantation with momentum $p$
ground state $\vert p>$ satisfying
\beq
 a_0 \vert p> =p\vert p>, \qquad    a_n\vert p>=0 \quad  n>0
\eeq
and we normal-order with respect to this vacuum.
 A state $\vert \Omega>$ is a highest weight state of $\W_{\infty/2}$ with
weights specified by an infinite-dimentional vector $(h^2, h^4, h^6
\ldots)$
if it satisfies
\bear
		  \W_n^{2r}\vert \Omega>&=&0,   \qquad  n>0 \nmb \\
		  (\W_0^{2r} -h^{2r})\vert \Omega>&=&0  \label{rimi}
\enar
for certain intercepts $h^{2r}$.

We next turn to the physical state conditions.
For any $\W $-string theory, a  general state is a linear combination of
states, each of which can be written as  a product of a matter-sector state $|
\psi>$ (an element of the $\phi$ Fock-space in our case) and a ghost-state. For
any given ghost-structure, the condition that the state represents a BRST
cohomology class is that $| \psi>$ should be a highest-weight state of the
appropriate $\W$-algebra
with  weights fixed  by  the choice of ghost structure.
 For example, for the bosonic
string with conventional ghost vacuum, physical states are highest-weight
representations of the  Virasoro algebra with weight (intercept)  $h^2=1$,
while for the
$\W_3$-string  with conventional ghost vacuum, the weights $(h^2, h^4)$ of
$(L_0, \W_0 )$ are $(4, 0)$
\cite{tm}.
For $\W$-algebras generated by a finite number of currents, there are a number
of other ways of determining the intercept for $L_0$ (see \cite{bg}), all of
which give the same answer.
For $\W$-strings generated by an infinite number of currents, such as
$\W_\infty$ and $\W_{\infty/2}$,   each method of calculating the weights of a
physical state gives a divergent sum, and different methods can give different
results, as can different ways of regularising a given sum.

Suppose a given $\W$-algebra is generated by a set of currents currents whose
spins $s$ form a set $\cal S$ (e.g. for $\W _ {\infty/2}$ we have ${\cal S}=
\left\{ 2,4,6, \dots \right\}$).
The corresponding $\W$-string has ghosts $\left\{ b^{s)},c^{(s)}|s\in{\cal S}
\right\}$ and the conventional ghost vacuum is given in terms of the
BRST-invariant vacuum $|0>$ by
\begin{equation}\label{state}
|\Omega>=\left( \prod_{s\in{\cal S}}{\prod_{i=1}^{s-1}{c_i^{(s)}}}
 \right) |0>
\end{equation}
The physical state conditions for a  state $|\psi> \otimes|\Omega>$ with this
ghost structure then imply that the matter sector state $|\psi>$ should be a
highest weight state of the $\W$-algebra with spin-two intercept  $h^2$ biven
by
\begin{equation}\label{intee}
h^2=\sum_ {s\in{\cal S}}\sum_{n=1}^{s-1}{n} =\sum_ {s\in{\cal S}}{1 \over
2}s(s-1)
\end{equation}
For the $\W_N$ string,  ${\cal S}= \left\{ 2,3,4,  \dots, N \right\}$ and
$h^2=1,4,\dots$ for $N=2,3, \dots$.
For the $\W_{\infty+1}$ string,
\begin{equation}\label{wtstr}
h^2= \sum_ {s=1}^ \infty{1 \over 2}s(s-1)= {1 \over 2}\left(
\zeta(-2)-\zeta(-1) \right)={1 \over 24}
\end{equation}
For the $\W_\infty$ string the intercept is the same sum but starting from
$s=2$ instead of $s=1$.
This can be evaluated as
  \begin{equation}\label{wtstra}
h^2= \sum_ {s=2}^ \infty{1 \over 2}s(s-1)=  \sum_ {s=1}^ \infty{1 \over
2}s(s-1) -0
 ={1 \over 24}
\end{equation}
or as
  \begin{equation}\label{wtstrb}
h^2= \sum_ {s=2}^ \infty{1 \over 2}s(s-1)=  \sum_ {r=1}^ \infty{1 \over
2}r(r+1)   = {1 \over 2}\left( \zeta(-2)+\zeta(-1) \right)=-{1 \over 24}
\end{equation}
Clearly, organising the series in different ways can give different results.
For $\W_{\infty/2}$, we obtain
\begin{equation}\label{wtstt}
h^2= \sum_ {s \ {\rm even}} {1 \over 2}s(s-1)
=  \sum_ {r=1}^ \infty r(2r-1)
= 2 \zeta(-2)-\zeta(-1) ={1 \over 12}
\end{equation}

Although it is reassuring that the calculation of the spin-two intercepts for
$\W_{1+\infty}$ and
$\W_{\infty/2}$ involved no rearrangement of the series, it is not clear
whether these are trustworthy results. Indeed, in \cite{pape}, the intercepts
for $\W_{1+\infty}$ and $\W_{ \infty}$
were calculated by a different method and different values were obtained
($h^2=0$ in both cases). Furthermore, choosing different ghost vacua will give
different intercepts and it is possible that for these theories there may be
physical states with   different
ghost structures.
As  there does not, at present,  seem to be a reliable way of fixing the
weights for infinite rank $\W$-algebras,    we shall proceed by analysing  the
one-boson highest weight representations of $\W _{\infty/2}$ for general
weights and  return to the
issue of how these weights are fixed in the $\W_{\infty/2}$-string at the end.

We shall now look for highest weight Fock-space states $\vert \Omega>$
satisfying (\ref{rimi}) which are eigenstates of momentum and level number
\beq
   a_0 \vert \Omega> =p\vert \Omega>, \qquad N\vert \Omega>= n \vert\Omega>
\eeq
for some momentum p, and level n where  $N=\sum a_n a_{-n}$
The mode expansion of $\W^{2s}$ gives the operators
\beq
   \W^{2s}_l=(-1)^{s-1}\sum_{k=1}^{2s-1} \Delta_k^{2s}
   \sum_{n=-\infty}^\infty n^{2s-k-1} (l-n)^{k-1}  :a_na_{l-n} :
\eeq
 where,
  \[ \Delta_k^{2s} = (-1)^k B(2s) A_k^{2s} \]
  and \[ B(2s) = \frac{ 2s^{2s-3} (2s)!}{(4s-3)!!} \]
       \[ A_k^{2s} ={1\over(2s-1)}{2s-1\choose k}{2s-1\choose 2s-k} \]
so that
\bear
   \W^{2s}_0 &=& 2(-1)^s \sum _{k=1}^{2s-1}B(2s)A^{2s}_k \sum_{n=1}^\infty
    a_{-n}a_n n^{2s-2} \nmb\\
	    &=&  2(-1)^s \sum _{k=1}^{2s-1}B(2s)A^{2s}_k
	    (a_{-1}a_1+ 4a_{-2}a_2 + 9a_{-3}a_3 + 16a_{-4}a_4+\ldots )
\enar

 Consider first highest weight states of level $0$, and momentum $p$.
 The constraint
\beq
(\W^2_0-h^2)\vert   p>=0
\eeq
implies that
\beq
p^2=-h^2,
\eeq
while the higher-spin
constraints are satisfied only if
\beq
h^4= h^6=h^8=\ldots       =0
\eeq
This is the case because for higher spin cases the operators
$\W_0^{2s}$ for $(s>1)$  do not contain any $a_0a_0$ term.
Thus states  of level $0$ only exist if
\[ h^4=h^6=h^8= \ldots       0     \]
with $h^2$ arbitrary.

Next consider states of level $1$, which are of the form
$\vert \Omega> = a_{-1}\vert p >$. The condition
\beq
\W^2_0 \vert \Omega> = h^2 \vert \Omega>
\eeq
implies
\beq
-(p^2+2) =h^2
\eeq
while $\W^2_1 \vert \Omega>=0$ implies
\[ p=0\]
Thus the only level-one  highest weight states must have zero momentum and
\beq
h^2 =-2
\eeq
Then
\beq
\W^4_1 \vert \Omega> = 0, \qquad     (\W^4_0 -h^4)\vert \Omega> = 0
\eeq
only if
\beq
h^4=2^5
\eeq

For higher-spin, the constraints (\ref{rimi}) imply
\bear
    h^ {2s} &=&2(-1)^{s-1} \sum_{k=1}^{2s-1} (-1)^{k-1}\Delta_k^{2s} \nmb\\
	    &=&2 (-1)^s \sum_{k=1}^{2s-1}\frac{ 2s^{2s-3} (2s)!}{(4s-3)!!}
    .{1\over(2s-1)}.{2s-1\choose k}{2s-1\choose 2s-k}\
\enar
For low s, this gives the results
\beq
h^6 = -2^9, h^8 = 2^{13}, h^{10} = -2^{17},..........
\eeq
and leads us to conjucture that the general result is
\beq
    h^{2s} =(-1)^s 2^{4s-3} \label{conj}
\eeq
Thus there are no highest weight states at level one unless the weights
$h^{2s}$ take certain fixed values, which we believe to be given by
(\ref{conj}).  If the weights do take these special values, then there is
  a physical level-one  state with zero momentum.

Now let us consider states of level $2$, of the form
    $ \vert \Omega>= \{ \alpha a_{-2}+ \beta ( a_{-1})^2 \}\vert p >$.
     This will be a highest weight state provided
\bear
  \W^{2s}_2 \vert \Omega > &=&0  \nmb\\
  \W^{2s}_1 \vert \Omega > &=&0  \label{babu} \\
  \W^{2s}_0 \vert \Omega > &=&h^{2s}  \vert \Omega > \nmb
\enar
     When $s=1$,   (\ref{babu}) implies
\bear
  \W^2_2 \vert \Omega>&=&0  \nmb\\
  \W^2_1 \vert \Omega>&=&0  \\
   \W^2_0 \vert \Omega>&=& h^2 \vert \Omega> \nmb
\enar
which is equivalent to
\bear
     -(a_1a_1 + 2a_0a_2 + \ldots )(\alpha a_{-2} + \beta (a_{-1})^2 )&=&0
\nmb\\
     -2(a_0a_1 + a_{-1}a_2 + \ldots )(\alpha a_{-2} + \beta (a_{-1})^2 )&=&0 \\
     -(a_0a_0 + 2a_{-1} a_1 +a_2a_{_2}+\ldots)
     (\alpha a_{-2} + \beta (a_{-1})^2 )
     &=& h^2  (\alpha a_{-2} + \beta (a_{-1})^2 ) \nmb
\enar
This implies
\bear
     2 p \alpha + \beta &=& 0  \\
     2 \alpha    + \beta p &=& 0 \\
     (p^2 + 4)            &=& h^2 \  \label{blob}
\enar
The
first two equations imply that either $\alpha=\beta=0$ in which case the state
is trivial or
\beq
    p^2= 1  \label{pisone}
\eeq
The final equation in (\ref{blob}) then fixes
\beq
    h^2=5
\eeq

 Similarly for the  $s=2$ components of (\ref{babu}), we obtain
\bear
    ( \beta - 8p \alpha) (\alpha a_{-2} + \beta a_{-1}^2)
&=& 0  \\
      (\beta p +22 \alpha)  (\alpha a_{-2} + \beta a_{-1}^2) &=& 0 \\
       -32(8 \alpha a_{-2} +\beta a_{-1}^2 )\vert p >
       &=& h^4  (\alpha a_{-2} + \beta a_{-1}^2 ) \
\enar
The first two equations imply that if the state is non-trivial, then
\beq
     p^2=- \frac {11}{4}
\eeq
This is inconsistent with (\ref{pisone}) and so we conclude that there are no
physical states at level two.

It is straightforward to give a similar analysis at higher levels and we have
checked that there are no physical states at levels 3, 4, or 5. In each case,
there are more constraints than there are free variables parameterizing the
state at that level, so that the system is over-constrained and there are no
solutions.  The number of contraints grows with level more rapidly than the
number of variables, and we conjecture that there are no further physical
states at any higher levels.

We now return to the $\W_{\infty/2}$-string. For a given ghost vacuum, the
physical state conditions imply that the matter part of a physical state should
be a highest weight state satisfying (\ref{rimi}) with weights
\begin{equation}\label{asdasd}
h^2= \bar h^2, \quad h^4= \bar h^4, \quad h^6=\bar h^6, \dots
\end{equation}
for some numbers $\bar h^{2r}$ which depend on the choice of ghost vacuum.
 Our results up to level five imply the following.
There is a physical state at level zero only if there is a ghost vacuum that
gives intercepts
\begin{equation}\label{lez}
 \bar h^4=\bar h^6=\bar h^8= \ldots     =  0
\end{equation}
with $\bar h^2 $ arbitrary. If so, then the physical state corresponds to a
particle in one dimension with momentum
$p= \sqrt{-\bar h ^2} $. Note that (\ref{wtstt}) suggests $\bar h^2= {1 \over
12}$ for the standard ghost vacuum.
At level one, there are no physical states unless the weights take fixed values
 \beq \label{leo}
\bar h^2=-2, \quad \bar h^4= 2^5, \quad \bar h^6 = -2^9, \quad \bar h^8 =
2^{13}, \quad \bar h^{10} = -2^{17},..........
\eeq
 in which case there is a single zero-momentum state.
For levels $2,3,4,5$ there can be no physical states at all, and we conjecture
that there are no states at higher levels either. If this is correct, then the
spectrum of this $\W_{\infty/2}$-string is very small indeed: it consists of at
most a massive particle, which will occur only if there is a ghost vacuum which
gives intercepts (\ref{lez}),  and a zero-momentum state, which will occur only
if there is a  ghost vacuum which gives intercepts (\ref{leo}).  However,
deciding whether or not these states actually occur requires a better
understanding of the new kind of divergencies that are present  in theories
with an infinite number of fields than we have at present.

\end{document}